# LEP asymmetries and fits of the Standard Model

B. Pietrzyk
Laboratoire de Physique des Particules LAPP, IN2P3-CNRS,
F-74941 Annecy-le-Vieux, France

**Abstract**

The lepton and quark asymmetries measured at LEP are presented. The results of the Standard Model fits to the electoweak data presented at this conference are given. The top mass obtained from the fit to the LEP data is $172^{+13+18}_{-14-20}$ GeV; it is $177^{+11+18}_{-11-19}$ when also the collider, $\nu$ and $A_{LR}$ data are included.

It is a well known feature of the Standard Model that the interactions at the $Z^0$ resonance at LEP depend on the spin orientation of the incoming electrons and positrons as well as outgoing fermions. This feature creates the difference between the forward $\sigma_F$ and backward $\sigma_B$ cross-sections measured by the forward-backward asymmetries

$$A_{FB} = \frac{\sigma_F - \sigma_B}{\sigma_F + \sigma_B}$$

LEP experiments have measured the lepton asymmetries at different center of mass energies. The slope of asymmetries as a function of energy is inversly proportional to axial vector coupling constant squared $g_A^2$ or to the leptonic width $\Gamma_l$. The measured asymmetries are extrapolated to $\sqrt{s} = M_Z$, then the $\gamma$-, $\gamma$Z-exchanges and the QED corrections are removed. We obtain in this way the $A^0 = \frac{3}{4} A_e A_f$ asymmetries, where

$$A_f = \frac{2 g_V g_A}{g_V^2 + g_A^2} = \frac{2(1 - 4Q \sin^2\theta_{\mathrm{w}}^{\mathrm{eff}})}{1 + (1 - 4Q \sin^2\theta_{\mathrm{w}}^{\mathrm{eff}})^2}$$

The couplings and $\sin^2\theta_{\mathrm{w}}^{\mathrm{eff}}$ are "effective" since in their definition all possible loops in the Z self energy and the vertex corrections are included.

The lepton asymmetries measured by the LEP experiments are presented in Fig. 1. Contrary to the other experiments the L3 result was not updated with the 93 data. The LEP average is 0.0170±0.0016 compared to 0.0161±0.0019 during the summer conferences[1]. The lepton universality of the asymmetries is checked on Fig. 1. In the total error of 0.0016 the common systematical error of 0.0008 due to uncertainty of LEP energy scale is included. Also the noncorrelated systematical experimental error (typicaly 0.0006) is included. The theoretical error is not included. The difference between two LEP fitting programs ZFITTER 4.6 and MIZA(BHM) is 0.0008. Half of this difference comes from the treatment of initial state radiation (O($\alpha^2$) vs.YFS). The rest of the difference is under investigation.

The $\tau$ polarisation is measured by the difference between right-handed $\sigma_R$ and left-handed $\sigma_L$ cross-section for the $\tau$ production.

$$P_\tau(cos\Theta) = \frac{\sigma_R - \sigma_L}{\sigma_R + \sigma_L} = -\frac{A_\tau(1 + cos^2\Theta) + 2A_e cos\Theta}{(1 + cos^2\Theta) + 2A_e A_\tau cos\Theta}$$

The mean $\tau$ polarisation measures $A_\tau$ while the forward-backward asymmetry of polarisation measures $A_e$. Typicaly the LEP experiments are measuring the angular distribution of the $\tau$ polarisation and obtain the $A_e$ and $A_\tau$ from the fit of this distribution.

The results of the $\tau$ polarisation and the $\tau$ polarisation asymmetries are presented in Fig. 1. The mean values of $A_e$ and $A_\tau$ from the $\tau$ polarisation give us $A_l$ assuming lepton universality. The value of $A_{LR}$ asymmetry from SLD [2], measuring also $A_e$, is given. There is a 1.9 $\sigma$ difference between $A_e$ and $A_\tau$ measurements at LEP, 2.4 $\sigma$ between $A_l$ and $A_{LR}$ and 3.1 $\sigma$ between $A_{LR}$ and $A_e$ from $\tau$ polarisation at LEP. The LEP data up to 1992 have been analysed, the error is dominated by statistics, so improvement can be expected with already existing data.

The results on b and c asymmetries and for different techniques of measurements are presented on Fig. 2. Also for these measurements the statistical error dominates and not all the collected data are yet analysed. The b and c asymmetries presented in Fig.2 are asymmetries at $\sqrt{s}$=91.3 GeV and are corrected for QCD effects. We obtain the $A_{FB}^{0,b} = 0.096 \pm 0.004(stat) \pm 0.002(sys)$ and $A_{FB}^{0,c} = 0.070 \pm 0.008 \pm 0.007$ after applying the energy shift to $\sqrt{s}$ = 91.19 GeV, introducing the QED radiative corrections, removing the photon exchange and the $\gamma Z^0$-interference.

The lower part of Fig. 2 shows the values of $\sin^2\theta_{\rm w}^{\rm eff}$ measured at LEP by different methods. The average value is 0.2322 ± 0.005 compared to 0.2321 ± 0.0006 reported during the 1993 summer conferences. It is interesting to note that the most precise measurement of $\sin^2\theta_{\rm w}^{\rm eff}$ comes from forward-backward b asymmetries.

The result of the recent left-right asymmetry experiment at SLC is also combined with the LEP results. The $\chi^2/(d.o.f.)$ of the combined average is increasing from 6.3/5 to 12.8/6 when the SLC result is included. The difference between the SLC result and the combined LEP number is 2.5 $\sigma$. The biggest difference, 3.1 $\sigma$, is between the SLC left-right asymmetry and the $\tau$ forward-backward asymmetries. Both are measuring $A_e$ and both errors are dominated by statistics.

The values of asymmetries as well as the other electroweak measurements presented at this conference are summarised in Table 1. They were used as input for the Standard Model fit. The results of the fit are presented in Table 2. A top mass of $172^{+13+18}_{-14-20}$ is obtained from the fit to the LEP results, with an error improved since the 1993 summer conferences value of $166^{+17+19}_{-19-22}$. The main improvement comes from the new Z width measurement. The central top mass value was obtained for the Higgs mass of 300 GeV and the second error describes the variation of the top mass with the Higgs mass between 60 and 1000 GeV. The $\alpha_s$ error of 0.005 becomes more and more interesting since the $\alpha_s$ measurement from $R_l$ is one of the cleanest from the theoretical point of view. The $\chi^2/(d.o.f.)$ is 11.4/9 in comparison to 3.5/8 during summer conferences. A top mass of $177^{+11+18}_{-11-19}$ is obtained when also the collider, $\nu$-N and $A_{LR}$ data are included in the fit. The $A_{LR}$ SLC value is responsible for the top mass rise in the fit as well as for the increase of $\chi^2/(d.o.f.)$ to 19.1/12. The Standard Model fit parameters resulting from the fit including all data for the Higgs mass of 300 GeV are presented in Table 1 as well as the pulls. The forward-backward $\tau$ polarization asymmetries and $\mathrm{R}_b$ are pushing the top mass down with pulls of 1.8 and 2.0 respectively while the $A_{LR}$ is driving it up with a pull of 2.7 .

The $\mathrm{R}_b = \Gamma_{\mathrm{b\bar{b}}}/\Gamma_{\mathrm{had}} = 0.2208 \pm 0.0024$ value is 2 $\sigma$ above the Standard Model prediction of 0.2158. The present value of $\mathrm{R}_b$ gives an upper limit on the top mass of 180 GeV independently of the Higgs mass. But if the difference of the $\mathrm{R}_b$ to the Standard Model persists it will have even more interesting consequences. It is interesting to follow this subject in the future. The $\mathrm{R}_b$ error of 0.0024 is obtained if the value of $\Gamma_{\mathrm{c\bar{c}}}$ is left free in the full heavy flavour electroweak fit . If the value of $R_{\rm c}$ is fixed to 0.171 then $\mathrm{R}_b$ is 0.2207 ± 0.0022. *

The comparison of the $\mathrm{R}_b$, $\sin^2\theta_{\rm w}^{\rm eff}$ and $R_\ell(R_{had})$ with the Standard Model predictions is presented in Figure 3. The deviation of $\mathrm{R}_b$ from the Standard Model can be seen on this plot. However $R_\ell$ is in agreement with the Standard Model for the $\alpha_s = 0.123 \pm 0.006$ used to produce the inclined band on this plot. For lower values of $\alpha_s$ the band moves up. It is seen from this plot that the present value of $\mathrm{R}_b$ favours the low top and low Higgs masses.

The results presented in this talk have been obtained by 4 LEP experiments and were averaged by the LEP Electroweak Working Group. I would like to thank D. Bardine, A. Blondel, J. Harton, S. de Jong, M. Martinez, M. Pepe-Altarelli, S. Rosier, D. Schaile, R. Tenchini and P. Wells for their help in preparation of this talk.

---

*In the Standard Model fit results presented here the value of $\mathrm{R}_b$ is slightly different from the value presented in my talk in Moriond due to some improvements in the averaging procedure. As a result the top mass is here slightly higher than in my Moriond presentation.

| | measurement | Standard Model fit | pull |
|---|---|---|---|
| a) LEP | | | |
| line-shape and lepton asymmetries: | | | |
| $M_Z$ [GeV] | $91.1895 \pm 0.0044$ | 91.192 | 0.6 |
| $\Gamma_Z$ [GeV] | $2.4969 \pm 0.0038$ | 2.4967 | 0.1 |
| $\sigma_h^0$ [nb] | $41.51 \pm 0.12$ | 41.44 | 0.6 |
| $R_\ell$ | $20.789 \pm 0.040$ | 20.781 | 0.2 |
| $A_{FB}^{0,\ell}$ | $0.0170 \pm 0.0016$ | 0.0152 | 1.1 |
| $\tau$ polarization: | | | |
| $\mathcal{A}_\tau$ | $0.150 \pm 0.010$ | 0.142 | 0.8 |
| $\mathcal{A}_e$ | $0.120 \pm 0.012$ | 0.142 | 1.8 |
| b and c quark results: | | | |
| $R_b = \Gamma_{b\bar{b}}/\Gamma_{had}$ | $0.2208 \pm 0.0024$ | 0.2158 | 2.0 |
| $R_c = \Gamma_{c\bar{c}}/\Gamma_{had}$ | $0.170 \pm 0.014$ | 0.172 | 0.1 |
| $A_{FB}^{0,b}$ | $0.0960 \pm 0.0043$ | 0.0997 | 0.8 |
| $A_{FB}^{0,c}$ | $0.070 \pm 0.011$ | 0.071 | 0.1 |
| $q\bar{q}$ charge asymmetry: | | | |
| $\sin^2\theta_{eff}^{lept}$ from $\langle Q_{FB} \rangle$ | $0.2320 \pm 0.0016$ | 0.2321 | 0.1 |
| b) $p\bar{p}$ and $\nu$N | | | |
| $M_W$ [GeV] (CDF, CDF prel., D0 prel., UA2; [3]) | $80.23 \pm 0.18$ | 80.31 | 0.4 |
| $1 - M_W^2/M_Z^2(\nu N)$ | $0.2256 \pm 0.0047$ | 0.2246 | 0.2 |
| c) SLC | | | |
| $\sin^2\theta_{eff}^{lept}$ from $\mathcal{A}_e$ | $0.2294 \pm 0.0010$ | 0.2321 | 2.7 |

Table 1: Summary of measurements included in the combined analysis of Standard Model parameters. Section a) summarizes LEP averages, section b) electroweak precision tests from hadron colliders and $\nu$N-scattering, section c) gives the result for $\sin^2\theta_{eff}^{lept}$ from the measurement of the left-right polarization asymmetry at SLC. The Standard Model fit result in column 3 and the pulls in column 4 are derived from the fit including all data (Table 2, column 4) for a fixed value of $M_H = 300$ GeV

| | LEP | LEP + Collider and $\nu$ data | LEP + Collider and $\nu$ data + $A_{\rm LR}$ from SLC |
|---|---|---|---|
| $M_{\rm t}$ (GeV) | $172^{+13}_{-14}\,^{+18}_{-20}$ | $170^{+12}_{-12}\,^{+18}_{-19}$ | $177^{+11}_{-11}\,^{+18}_{-19}$ |
| $\alpha_s(M_{\rm Z}^2)$ | $0.125 \pm 0.005\ \pm 0.002$ | $0.125 \pm 0.005\ \pm 0.002$ | $0.124 \pm 0.005\ \pm 0.002$ |
| $\chi^2/(d.o.f.)$ | 11.4/9 | 11.5/11 | 19.1/12 |
| $\sin^2\theta_{eff}^{\rm lept}$ | $0.2323 \pm 0.0002\ ^{+0.0001}_{-0.0002}$ | $0.2324 \pm 0.0002\ ^{+0.0001}_{-0.0002}$ | $0.2320 \pm 0.0003\ ^{+0.0001}_{-0.0002}$ |
| $1 - M_{\rm W}^2/M_{\rm Z}^2$ | $0.2251 \pm 0.0015\ ^{+0.0003}_{-0.0003}$ | $0.2253 \pm 0.0013\ ^{+0.0003}_{-0.0002}$ | $0.2243 \pm 0.0012\ ^{+0.0003}_{-0.0002}$ |
| $M_{\rm W}$ (GeV) | $80.28 \pm 0.08\ ^{+0.01}_{-0.02}$ | $80.26 \pm 0.07\ ^{+0.01}_{-0.01}$ | $80.31 \pm 0.06\ ^{+0.01}_{-0.01}$ |

Table 2: Results of fits to LEP and other data for $M_{\rm t}$ and $\alpha_s(M_{\rm Z}^2)$. No external constraint on $\alpha_s(M_{\rm Z}^2)$ has been imposed. In the third column also the combined data from the $\rm p\overline{p}$ experiments UA2 [4], CDF [5, 6] and D0 [6]: $M_{\rm W} = 80.22 \pm 0.16$ GeV and from the neutrino experiments, CDHS [7], CHARM [8] and CCFR [9]: $1 - M_{\rm W}^2/M_{\rm Z}^2 = 0.2256 \pm 0.0047$ are included. The fourth column gives the result when also the SLD measurement of the left-right asymmetry at SLC [2], $\sin^2\theta_{eff}^{\rm lept} = 0.2294 \pm 0.0010$, is added. The central values and the first errors quoted refer to $M_{\rm H} = 300$ GeV. The second errors correspond to the variation of the central value when varying $M_{\rm H}$ in the interval $60 \leq M_{\rm H}$ [GeV] $\leq 1000$

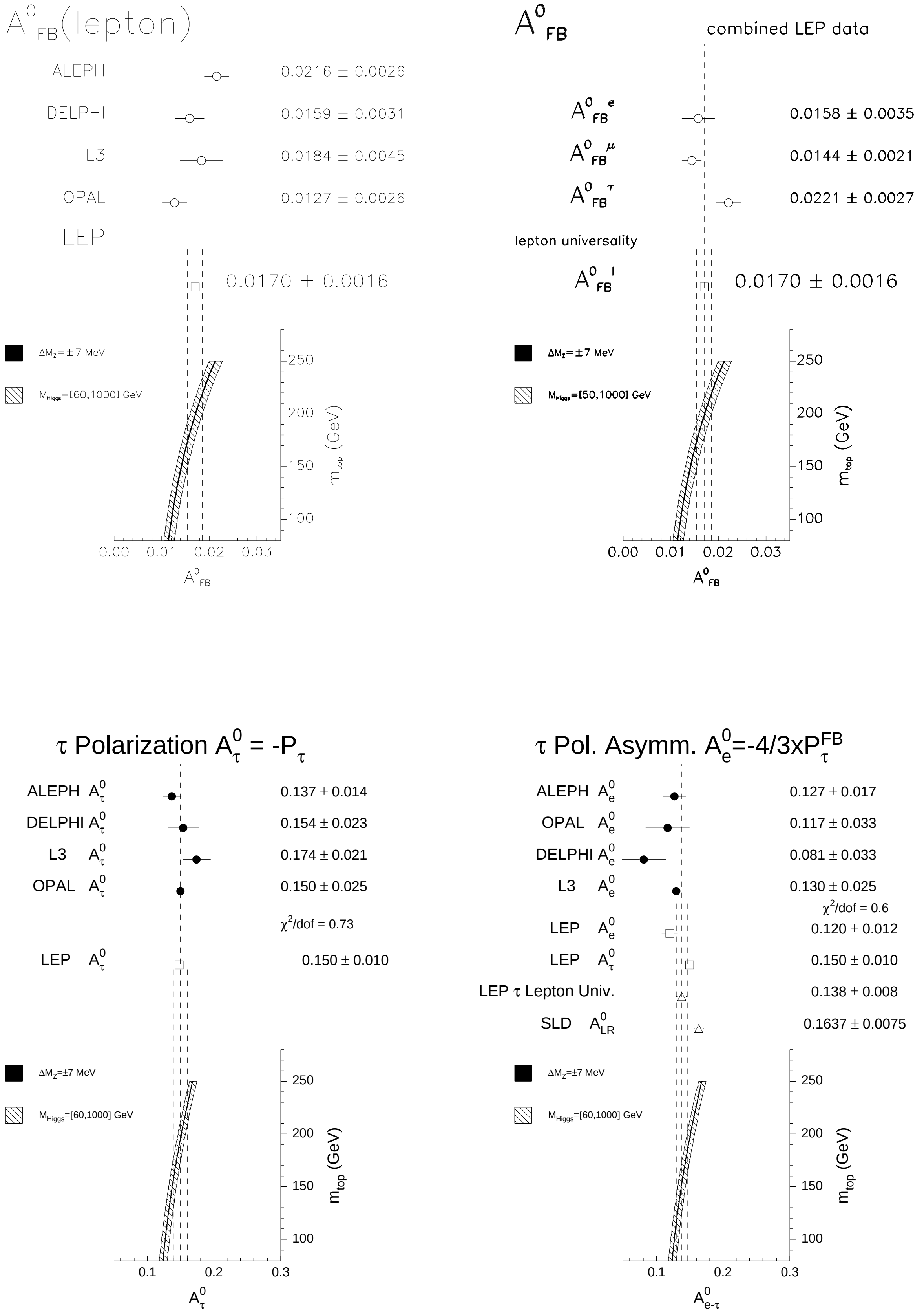

Figure 1: Forward-backward lepton asymmetries, $\tau$ polarization and $\tau$ polarization asymmetries from LEP data.

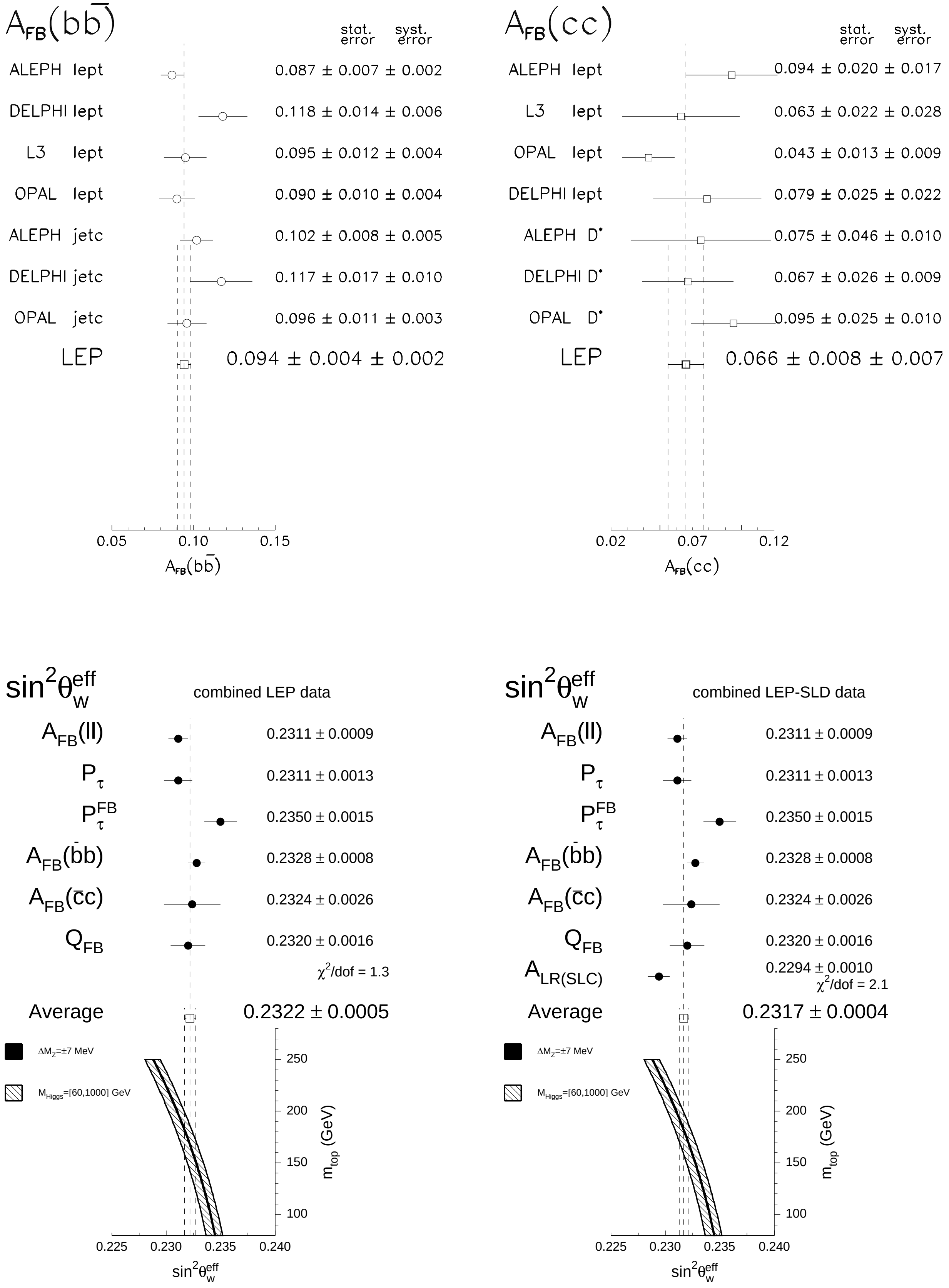

Figure 2: Forward-backward b abd c asymmetries and $\sin^2\theta_w^{\rm eff}$ measurements from LEP and SLC data.

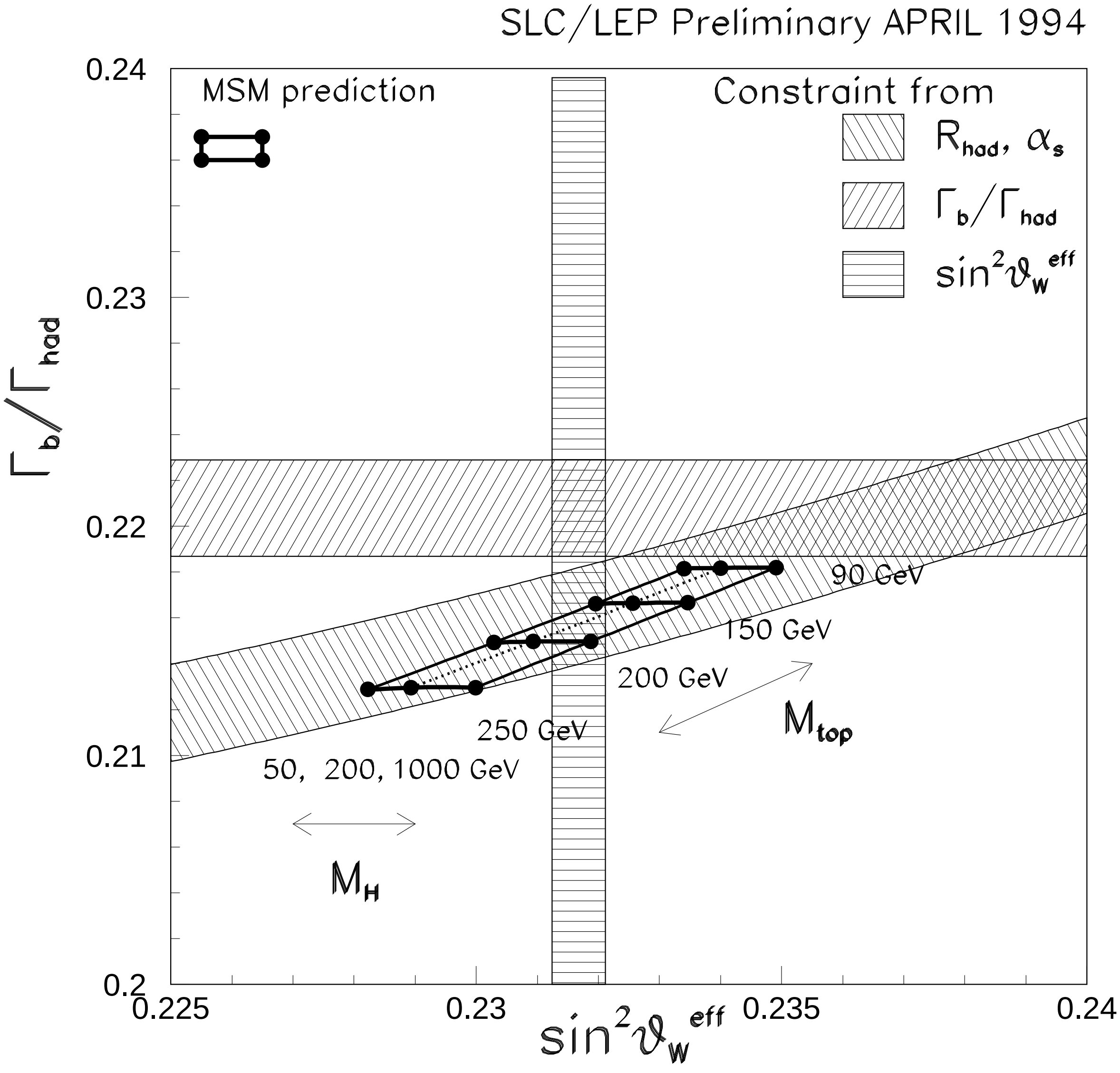

Figure 3: The LEP/SLC measurements of $\sin^2\theta_w^{eff}$ and $R_b$ and the Standard Model predictions (updated figure from ref. [10]).